\documentclass{ifacconf}

\usepackage{graphicx}      
\usepackage{url}
\usepackage{array}   
\usepackage{booktabs} 
\usepackage{caption} 
\usepackage{siunitx} 
\usepackage{epstopdf} 
\usepackage{xcolor}
\usepackage[normalem]{ulem}

\usepackage{subcaption}
\usepackage{tabularx}
\usepackage{amsmath,amssymb,amsfonts,mathtools,nccmath,bm, tikz, enumerate}
\usepackage{multicol}
\newtheorem{definition}{Definition}[section] 
\newtheorem{assumption}{Assumption}[section]
\newtheorem{remark}{Remark}[section]
\newtheorem{theorem}{Theorem}[section]
\newtheorem{lemma}[theorem]{Lemma}
\usepackage{algorithmic}
\usepackage{graphicx}
\usepackage{textcomp}
\usepackage{cuted}
\usepackage{float}
\usepackage{xcolor}
\def\BibTeX{{\rm B\kern-.05em{\sc i\kern-.025em b}\kern-.08em
    T\kern-.1667em\lower.7ex\hbox{E}\kern-.125emX}}
\allowdisplaybreaks

\usepackage{tabstackengine}
\usetikzlibrary{shapes,shapes.geometric,arrows.meta,fit,calc,positioning,automata,patterns}
\usetikzlibrary{tikzmark,calc,decorations.pathreplacing, backgrounds}
\tikzset{
    node_style/.style={
        draw, fill=red!30, text=black, regular polygon, regular polygon sides=8,
        minimum size=0.4cm, font=\scriptsize\bfseries
    },
    edge_style/.style={draw, ->, thick, purple},
    edge_style_1/.style={draw, <->, thick, purple},
}

\usepackage{natbib}        
\begin{document}
\begin{frontmatter}

\title{Robust Multi-Agent Safety via Tube-Based Tightened Exponential Barrier Functions} 


\author[First]{Armel Koulong} 
\author[First]{Ali Pakniyat} 

\address[First]{$\mathcal{M}^{\,3\!}\mathcal{A\,C}$ (Multi-Modal Multi-Agent Control) Lab, \\ University of Alabama, Tuscaloosa, AL, USA (e-mail: akoulongzoyem@crimson.ua.edu, apakniyat@ua.edu).}

\begin{abstract}                
This paper presents a constructive framework for synthesizing provably safe controllers for nonlinear multi-agent systems subject to bounded disturbances. The methodology applies to systems representable in Brunovsky canonical form, accommodating arbitrary-order dynamics in multi-dimensional spaces. The central contribution is a method of constraint tightening that formally couples robust error feedback with nominal trajectory planning. The key insight is that the design of an ancillary feedback law, which confines state errors to a robust positively invariant (RPI) tube, simultaneously provides the exact information needed to ensure the safety of the nominal plan. Specifically, the geometry of the resulting RPI tube is leveraged via its support function to derive state-dependent safety margins. These margins are then used to systematically tighten the high relative-degree exponential control barrier function (eCBF) constraints imposed on the nominal planner. This integrated synthesis guarantees that any nominal trajectory satisfying the tightened constraints corresponds to a provably safe trajectory for the true, disturbed system. We demonstrate the practical utility of this formal synthesis method by implementing the planner within a distributed Model Predictive Control (MPC) scheme, which optimizes performance while inheriting the robust safety guarantees.
\end{abstract}


\end{frontmatter}

\section{Introduction} \label{Introduction}


To operate successfully in applications like vehicle convoys and aerial swarms, multi-agent systems (MAS) must satisfy strict coordination and safety requirements \citep{Barnes2007}. This challenge becomes more severe for systems with high relative degrees, where chains of integrators between the control input and safety output slow the effective response and increase collision risk in safety-critical settings \citep{Nguyen2016,GARG2024}. Model Predictive Control (MPC) provides a powerful framework for constrained control by optimizing over predicted future trajectories \citep{mayne2000constrained}, while higher-order and exponential control barrier functions (HOCBFs and eCBFs) enable systematic enforcement of safety constraints in systems with complex, high-relative-degree dynamics \citep{Wei2021,Wei2019,Xiao2021,Nguyen2016,Wang2021}. However, guaranteeing robust safety for MAS under model uncertainty and bounded disturbances remains challenging, particularly when these advanced CBF techniques are deployed in distributed MPC architectures.

This work introduces a robust safety framework for leader–follower formation control in continuous-time, nonlinear multi-agent systems subject to bounded matched disturbances. The central contribution is a method that formally couples robust error feedback with nominal trajectory planning to enforce safety specifications. Our approach complements existing distributed MPC–HOCBF integrations \citep{wang2025}, whose safety analysis is often confined to nominal, disturbance-free models, and extends beyond other distributed controllers for multi-agent systems that are either reactive and lack the foresight of a prediction horizon \citep{Tan2022,Mestres2024} or, when predictive, do not address robustness for high-relative-degree dynamics \citep{Jiang2024}. By integrating tube-based robust constructions \citep{Kolathaya2019} with techniques for high-order constraints \citep{Wei2019,Ames2017}, the proposed framework provides a unified setting that combines predictive planning, robustness to bounded disturbances, and applicability to high-relative-degree multi-agent systems.

Safety-constrained control is often formulated either via barrier Lyapunov functions (BLFs), which enforce constraints through Lyapunov-like functions that diverge at the boundary, e.g., \citep{koulong2025acc, Zhao2025}, or via control barrier functions (CBFs), which encode forward invariance as affine-in-control inequalities that integrate naturally with optimization-based controllers such as QP filters and MPC. In this work, we adopt high-relative-degree CBF variants (HOCBF/eCBF) for this tractability and focus on robustness by converting tube-bounded tracking errors into systematic tightenings of the eCBF constraints, so that nominal planning implies safety for the true disturbed system. This perspective is related to robust CBF ideas such as the ``tube-CBF'' construction of \citep{Schilliger2021}, as well as to single-agent plan–track and safety-filter frameworks including FaSTrack \citep{Herbert2017}, safety filters \citep{Hsu2024}, and funnel libraries \citep{Anirudha2017}. For multi-agent systems, by contrast, the dominant safety constraints are mutual collision-avoidance conditions of the form $h_{ij}(x_i,x_j) \geq 0$, which couple agents’ states and disturbance tubes and cannot be enforced by applying independent copies of a single-agent tube, funnel, or safety filter to each agent. Extending these ideas to multi-agent settings is therefore  nontrivial, 
and robustness must be propagated through joint constraints involving multiple agents’ uncertainty sets; our key contribution is a support-function-based tightening that aggregates per-agent tubes into explicit scalar safety margins for pairwise constraints, enabling distributed MPC feasibility to imply robust safety for continuous-time nonlinear multi-agent systems with high-relative-degree safety outputs.

We propose a control synthesis framework that couples ancillary robust feedback with nominal planning. Each agent uses a local feedback law to confine nominal–true deviations within a precomputed RPI tube, and the tube geometry is then used to systematically tighten the eCBF constraints enforced on the nominal trajectories, so that nominal feasibility implies safety of the true disturbed system. We embed these tightened eCBF constraints in a distributed MPC formulation, where the prediction horizon enables proactive collision/obstacle avoidance and yields a tractable local optimization with end-to-end guarantees of robust safety and recursive feasibility.

The remainder of the paper is organized as follows: Section~\ref{Methodology} introduces the dynamics, nominal model, RPI tubes, and support-function-based eCBF tightening; Section~\ref{sec:MainResult} presents the DMPC formulation with tightened eCBF constraints under a fixed communication topology; Section~\ref{NUMERICALEXAMPLE} provides a numerical validation; and Section~\ref{CONCLUSION} concludes with contributions and future directions.

\section{Problem Formulation and Methodology}\label{Methodology}

We consider a system composed of $N$ follower agents indexed by $i\in\mathcal{V}=\{1,\dots,N\}$ and a leader agent indexed~by~$0$. The dynamics of the \(i^\text{th}\) follower agent is described by the $n$-th order nonlinear Brunovsky form as:
\begin{align}\label{eq:ct_follower}
\dot{x}_p^{\,i} &= x_{p+1}^{\,i}, &p =1,\dots,n-1, \notag\\
\dot{x}_n^{\,i} &= f^{\,i}(x^{\,i},t) + u^{\,i} + w^{\,i}, &w^{\,i}(t)\in\mathcal D_i.
\end{align} 
where $x_p^i\in\mathbb{R}^d$ is the $p$-th component of the state of agent $i$ with dimension $d$, $u^i\in\mathbb{R}^d$ is its input, and $w^i\in\mathbb{R}^d$ is a bounded time-varying disturbance for agent $i$.
The dynamics of the leader agent is given as:
\begin{align}\label{eq:ct_leader}
\dot{x}_p^{\,0} &= x_{p+1}^{\,0},\hspace{85pt} p =1,\dots,n-1, \notag\\
\dot{x}_n^{\,0} &= f^{\,0}(x^{\,0},t).
\end{align}

To manage the disturbance $w^{i}(t)$ in \eqref{eq:ct_follower}, our strategy is to design the high-level planner for a simplified nominal system and use a local feedback controller to confine the resulting error dynamics within a robust positively invariant (RPI) tube \citep{rawlings2017model,MAYNE200736}. 
We therefore define a nominal trajectory $(\bar{x}^i,\bar{u}^i)$ for \eqref{eq:ct_follower} and its corresponding nominal system as
\begin{align}\label{eq:ct_follower_nominal}
\dot{\bar x}_p^{\,i} &= \bar x_{p+1}^{\,i}, \hspace{85pt}p=1,\dots,n-1, \notag\\
\dot{\bar x}_n^{\,i} &= f^{\,i}(\bar x^{\,i},t) + \bar u^{\,i},
\end{align}
with $\bar x^{\,i}:=[\bar x_1^{\,i};\dots;\bar x_n^{\,i}]$, and define the nominal error as $
z_p^{\,i}:=x_p^{\,i}-\bar x_p^{\,i},\quad p=1,\dots,n
$, which is equivalent to defining $
z^{\,i} :=x^{\,i}-\bar x^{\,i} =[z_1^{\,i};\dots;z_n^{\,i}]\in\mathbb{R}^{nd}.
$
The nominal error dynamics is therefore
\begin{align}\label{eq:nominal_error_dynamics_block}
\dot z^i_p &= z^i_{p+1}\quad (p=1,\dots,n-1), \notag \\
\dot z^i_n &= \Delta f^i(z^i,\bar x^i,t) + (u^i-\bar u^i) + w^i,
\end{align}
where $\Delta f^i(z^i,\bar x^i,t):=f^i(\bar x^i+z^i,t)-f^i(\bar x^i,t)$.
Defining 
\begin{equation} \label{eq:AGdefinition}
A_{0}^{i}:=\begin{bmatrix}0 & I_{d} & \cdots & 0\\
\vdots & \vdots & \ddots & \vdots\\
0 & 0 & \cdots & I_{d}\\
0 & 0 & \cdots & 0
\end{bmatrix}, \quad G^{i} := \begin{bmatrix}0\\\vdots\\0\\ I_d \end{bmatrix},    
\end{equation}
we write the error dynamics \eqref{eq:nominal_error_dynamics_block} as
\begin{align}\label{eq:stacked_nominal_error_dyn}
\dot z^{\,i} \;=\; A_0^{\,i}\,z^{\,i} \;+\; G^{\,i}\big(\Delta f^{\,i} + (u^{\,i}-\bar u^{\,i}) + w^{\,i}\big).
\end{align}

\subsection{State Based Ancillary Feedback:}

Given the error $z^{\,i} = x^{\,i} - \bar{x}^{\,i}$, the ancillary tube feedback is designed~as
\begin{align}\label{eq:ancillary_control}
u^{\,i} &=\bar u^{\,i} -\sum_{p=1}^n K_p^{\,i}\,z_p^{\,i} \equiv \bar u^{\,i}-K^{\,i}z^{\,i}, 
\end{align}
where $K_p^{\,i}\in\mathbb{R}^{d\times d}$ are constant ancillary gains (chosen to stabilize the chain of integrators) and $K^{\,i}$ is the corresponding matrix representation of the gains. 
Substituting \eqref{eq:ancillary_control} into \eqref{eq:stacked_nominal_error_dyn}
gives 
\begin{equation}\label{eq:stacked_nominal_error_dyn_upd}
    \dot z^{\,i} \;=\; A_K^{\,i}\,z^{\,i} \;+\; G^{\,i}\big(\Delta f^{\,i}(z^i,\bar{x}^i,t) + w^{\,i}\big),
\end{equation}
where $A_K^{\,i}:=A_0^{\,i}-G^{\,i}K^{\,i}$, by construction \eqref{eq:AGdefinition}, is a constant matrix. The gains $K_p^{\,i}$ must be picked so that $A_K^{\,i}$ is Hurwitz. This can be achieved noting that for any $Q_i\succ0$, the continuous-time Lyapunov equation is:
\begin{equation}\label{eq:Lyapunov_eq_1}
(A_K^{\,i})^\top P_i + P_i A_K^{\,i} = -Q_i,\qquad P_i\succ0.
\end{equation}


\subsection{Robust
Positive Invariant (RPI) Tube Construction:}\label{par:tube_inv}

\begin{definition}[Robust
Positive Invariance (RPI)]
A set $\mathcal{Z}_i \subset \mathbb{R}^n$ is said to be robustly
positively invariant if, for all $z^i(t_0) \in \mathcal{Z}_i$ and any $w^i(t) \in \mathcal D_i$,  the condition $z^i(t) \in \mathcal{Z}_i$ holds for all $t \geq t_0$ \citep{Blanchini2007}. 
\hfill $\square$
\end{definition}

\begin{lemma}[RPI ellipsoid]\label{lem:RPI_tube}
Let $\|w^{\,i}(t)\|\le \bar w_i$ for all $t \geq t_0$ and that $Q_i$ and the corresponding $P_i$ in \eqref{eq:Lyapunov_eq_1} are such that, \mbox{$\frac{\lambda_{\min}(Q_{i})}{2\lambda_{\max}(P_{i})} > L_{f^i}$}. 
Then the per-agent ellipsoidal tube
\begin{align}
\label{eq:Tube}
\mathcal Z_i = \{\,z^{\,i}:~ (z^{\,i})^\top P_i z^{\,i}\le \rho_i^{\,2}\,\}
\end{align}
is robust positively invariant 
for all $\rho_i$ satisfying
\begin{align*}
 \rho_i \geq \frac{2\bar{w}_{i}\lambda_{\max}(P_{i})}{\lambda_{\min}(Q_{i})-2L_{f^{i}}\lambda_{\max}(P_{i})},
\end{align*}
where $L_{f^i}$ is an upper bound on the Lipchitz constant of~$f^i$ on a neighborhood of $\bar{x}^i$ containing the $\mathcal{Z}_i$ variation set.
\hfill $\square$
\end{lemma}

\begin{pf}
For the Lyapunov function $V_i(z^{\,i}) := \|z^{i}\|_{P_i}^2 \equiv (z^{\,i})^\top P_i z^{\,i}$, we have
\begin{multline} \label{eq:neg_lyap}
\hspace{-9pt} \frac{d}{dt}\,V_i(z^{\,i}(t))
= \frac{d}{dt}\big((z^{\,i})^\top P_i z^{\,i}\big)
\\
= {\dot z^{\,i\,\top} P_i z^{\,i}} + {(z^{\,i})^\top P_i \dot z^{\,i}} =\; 2\,(z^{\,i})^\top P_i\,\dot z^{\,i} \hfill
\\
\overset{\eqref{eq:stacked_nominal_error_dyn_upd}}{=} 2 (z^{\,i})^\top P_i A_K^{\,i} z^{\,i}
+ 2 (z^{\,i})^\top P_i G^{\,i}\,\Delta f^{\,i}
+ 2 (z^{\,i})^\top P_i G^{\,i}\,w^{\,i} \hfill
\\
= (z^{\,i})^\top\!\big(P_i A_K^{\,i} + (A_K^{\,i})^\top P_i\big) z^{\,i} \hfill
\\ \hfill + 2 (z^{\,i})^\top P_i G^{\,i}\,\Delta f^{\,i}
+ 2 (z^{\,i})^\top P_i G^{\,i}\,w^{\,i}
\\
\overset{\eqref{eq:Lyapunov_eq_1}}{=}- (z^{i})^\top Q_i z^{i}
+ 2 (z^{i})^\top P_i G^{i}\Delta f^{i}
+ 2 (z^{i})^\top P_i G^{i}w^{i}.
\end{multline}

The Rayleigh-quotient bound on $Q_i$ is written as
\begin{equation} \label{eq:Rayleigh}
    -(z^{i})^\top Q_i z^{i} \leq -\lambda_{\min}(Q_i) \|z^{i}\|^2
\end{equation}
Also, by the Cauchy–Schwarz inequality
\begin{multline} \label{eq:CauchySchwarzF}
    2 (z^{i})^\top P_i G^{i} \Delta f^{i}
    \leq 2 \|z^{i}\| \cdot \|P_i\| \cdot \|G^{i} \Delta f^{i}\|
    \\
    \leq 2 L_{f^i} \lambda_{\max}(P_i) \|z^{i}\|^2,
\end{multline}
and, similarly, 
\begin{multline}\label{eq:CauchySchwarzW}
2 (z^{i})^\top P_i G^{i} w^{i} \leq 2 \|z^{i}\| \cdot \|P_i\| \cdot \|G^{i} \Delta w^{i}\|
\\
\leq 2 \bar w_i \lambda_{\max}(P_i) \|z^{i}\|.
\end{multline}
Employing \eqref{eq:Rayleigh}, \eqref{eq:CauchySchwarzF} and \eqref{eq:CauchySchwarzW}, we obtain from \eqref{eq:neg_lyap} that
\begin{multline} \label{eq:negV_intermediate}
    \frac{d}{dt} V_i(z^{i}) \leq -\big(\lambda_{\min}(Q_i) - 2 L_{f^i} \lambda_{\max}(P_i) \big) \|z^{i}\|^2
    \\
    + 2 \bar w_i \lambda_{\max}(P_i) \|z^{i}\|.
\end{multline}
For the set $\mathcal{Z}_i$ to be positively invariant, its is required that $\frac{d}{dt}V_i(z^i) \leq 0$ for any state $z^i$ on the boundary $\partial \mathcal{Z}_i$ of the tube, which is identified by  $V_i(z^i) = \rho_i^2$. This requires that
\begin{equation}
    -\big(\lambda_{\min}(Q_i) - 2 L_{f^i} \lambda_{\max}(P_i) \big) \rho_i^2
    + 2 \bar w_i \lambda_{\max}(P_i) \rho_i \leq 0.
\end{equation}
it suffices that
\begin{equation}
\rho_{i}\geq\frac{2\bar{w}_{i}\lambda_{\max}(P_{i})}{\lambda_{\min}(Q_{i})-2L_{f^{i}}\lambda_{\max}(P_{i})}
\end{equation}
\hfill $\blacksquare$
\end{pf}

We remark that the tube radius $\rho_i$ depends on a local Lipschitz constant $L_{f^i}$ of $f^i$ over a compact region containing the tube $Z_i$, and this constant in turn depends on the size of that region. In practice, one can start from a conservative upper bound on $L_{f^i}$, compute the resulting tube $Z_i$ and radius $\rho_i$, and then iteratively refine $L_{f^i}$ and $\rho_i$ over the compact operating region thus obtained.

\subsection{Tightened Safety and Support Functions}
To enforce safety for the true state $x^{i}$ while optimizing over the nominal state $\bar x^{i}$, we derive a tighten term using the per-agent RPI tubes $\mathcal{Z}i$ from Lemma~\ref{lem:RPI_tube}. This is accomplished by employing the support function of the ellipsoidal tube \citep{Arcari2023,VILLANUEVA2017}, which calculates the maximum extent of the tube in a specified direction and projects the RPI tube onto the gradient of the eCBF, converting the set of possible disturbances into a single scalar representing the worst-case deviation along the most critical direction. This scalar offset is later used in Section~\ref{sec:MainResult} to tighten the nominal eCBF inequality, ensuring that even under worst-case disturbance, the true state remains~safe. 

\begin{assumption}\label{assum:ConvexSafetyFunctions}
For each pair $(i,j) \in \mathcal{V} \times \mathcal{V}$ and each obstacle $O \in \mathcal{O}$, there exist scalar safety functions
\mbox{$h_{ij}:\mathbb{R}^{n \cdot d}\times\mathbb{R}^{n \cdot d}\to\mathbb{R}$} and
$h_{iO}:\mathbb{R}^{n \cdot d}\to\mathbb{R}$ that encode safety via $h_{ij}(x^i,x^j)\ge 0$ and $h_{iO}(x^i)\ge 0$.
We assume $h_{ij}$ and $h_{iO}$ are convex and continuously differentiable in their arguments.
\hfill $\square$
\end{assumption}

To facilitate the discussion, the control-affine dynamics of agent $i$ from \eqref{eq:ct_follower} is written
\begin{align}\label{eq:ct_follower_affine}
    \dot{{x}}^i = F_{x}({x}^i) + F_{u} {u}^i + F_{w} w^i
\end{align}
where
$
F_{x}(x^{i})=\big[\begin{array}{ccc}
{x}_{2}^{i^{\top}} & \cdots & f^{i}(x^{i},t)^{\top}\end{array}\big]^{\top},$
$
F_{u}=[\begin{array}{cccc}
0 & \cdots & 0 & I\end{array}]^{\top},
$ $F_{w}=[\begin{array}{cccc}
0 & \cdots & 0 & I\end{array}]^{\top}$. Accordingly, the nominal control–affine dynamics of agent~$i$ from \eqref{eq:ct_follower_nominal} is written 
\begin{align}\label{eq:ct_follower_nominal_affine}
    \dot{\bar{x}}^i = F_{x}(\bar{x}^i) + F_{u} \bar{u}^i
\end{align}

Given a scalar function $h_{\bullet}: \mathbb{R}^n \to \mathbb{R}$, 
the $r$-th order Lie derivatives of $h_{\bullet}$ along $F_{x}$ are defined recursively as:
\begin{align*}
L_{F_{x}}^{0} h_{\bullet} &:= h_{\bullet}, 
\\
L_{F_{x}}^{1} h_{\bullet} &:= \nabla h_{\bullet}^\top F_{x}, \\
L_{F_{x}}^{r} h_{\bullet} &:= \nabla(L_{F_{x}}^{r-1} h_{\bullet})^\top F_{x}, \quad r \geq 2.
\end{align*}
Similarly, the Lie derivatives of $h_{\bullet}$ along $F_{u}$ is defined as:
\begin{align*}
L_{F_{u}} h_{\bullet} := \nabla h_{\bullet}^\top F_{u}.
\end{align*}

\begin{definition}[Relative degree $r$] \label{def:RelDeg}
    The relative degree $r$ of the function $h_{\bullet}$ with respect to the input whenever $L_{F_u} L_{F_x}^{j} h = 0$ for $j=1,\ldots,r-2$ and $L_{F_u} L_{F_x}^{r-1} h \neq 0$
    \citep{Nguyen2016}. 
    \hfill $\square$
\end{definition}

\begin{definition}(\textit{eCBF}\label{def:true_safety_sets} \citep[Definition~5.1]{nguyen2017robust})
For the system \eqref{eq:ct_follower_affine}, and the safety set
\begin{multline} \label{eq:SafetySet}
    \mathcal{C}:=\bigg\{ \left(x^{1},\cdots,x^{i},\cdots,x^{N}\right)\in\mathbb{R}^{N\cdot n\cdot d}:
    \\[-6pt]
    \hfill h_{ij}(x^{i},x^{j})\geq0, \hspace{20pt} \forall(i,j)\in\mathcal{V}^{2}, \hspace{45pt}
    \\[-6pt]
    h_{iO}(x^{i})\geq0, \hspace{20pt}\forall(i,O)\in\mathcal{V}\times\mathcal{O}_{i}\bigg\} ,
\end{multline}
the functions $h_{ij}(x^{i},x^{j})$ (similarly $h_{iO}(x^{i})$) are exponential control barrier functions (eCBF) of relative degree $r$, if there exist $K_b^{ij} \in \mathbb{R}^{r \times 1}$ s.t.,
\begin{equation}\label{eq:ecbf_ds}
    \inf_{u^i \in {U}_i} \left( L_{F_{x}}^r h_{ij} + L_{F_u} L_{F_{x}}^{r-1} h_{ij} \cdot u^i + K_b^{ij} \eta_{ij} \right) \ge 0, 
\end{equation}
with $\eta_{ij}(x^{i}(t),x^{j}(t)) = \begin{bmatrix} h_{ij} , L_{F_{x}} h_{ij} , \ldots , L_{F_{x}}^{r-1} h_{ij} \end{bmatrix}^\top$, for all $(i,j)\in\mathcal{V}^{2}$ and $(\cdot, \cdots, x^i, \cdots, x^j,\cdots, \cdot) \in \mathcal{C}$, and 
\begin{align}
h_{ij}(x^{i}(t),x^{j}(t)) \ge C_b\,e^{{A_b^{ij}}t}\eta_{ij}(x^{i}(t_0),x^{j}(t_0)) \ge 0, \; 
\end{align}
whenever \(h_{ij}(x^{i}(t_0),x^{j}(t_0)) \ge 0,\)
where the matrix $A_b^{ij}$ depends on the choice of $K_b^{ij}$, and \mbox{$C_b = [1, 0, \ldots, 0]$}. 
\hfill $\square$
\end{definition}


\begin{definition}[Tightened Safety Function]\label{def:tightened_safety_fxn}
Given the decomposition $x^{\,i}=\bar x^{\,i}+z^{\,i}$, $x^{\,j}=\bar x^{\,j}+z^{\,j}$ with $z^{\,i}\in\mathcal Z_i$, $z^{\,j}\in\mathcal Z_j$, and $z^{ij}=z^i - z^j$, 
we define the inter-agent tightened safety function as the worst-case value (infimum) considering all possible errors in relative tube $\mathcal Z_{ij}=\mathcal Z_i\oplus(-\mathcal Z_j)$ in the form of (illustrated in Figure~\ref{fig:tube_tightening})
\begin{align}\label{eq:tightened_fxn}
h_{ij}^{\text{tight}}(\bar x^{\,i},\bar x^{\,j}) &:= \inf_{z^{ij} \in \mathcal{Z}_{ij}} h_{ij}(x^{\,i},x^{\,j}) \notag \\ &\equiv \inf_{z^{ij} \in \mathcal{Z}_{ij}} h_{ij}(\bar x^{\,i}+z^{\,i},\bar x^{\,j}+z^{\,j}),
\end{align}
which occurs at the tube configuration where inter-agent  $i\text{--}j$ and agent-obstacle $i\text{--}O$ are close to violating safety. 
Similarly, for $i\text{--}O$, $h_{iO}^{\text{tight}}(\bar{x}^{i}) 
:= \inf_{z^{i} \in \mathcal{Z}_i} h_{iO}(\bar{x}^{i} + z^{i}).$
\hfill $\square$
\end{definition}

\begin{definition}[Support Function of RPI Tubes]\label{def:support_fxn}
The support function of $\mathcal {Z}_i$ is defined  \citep{boyd2004convex} as
\begin{align}
\sigma_{\mathcal {Z}_i}(g_i) := \sup_{z^{\,i}\in\mathcal {Z}_i} g_i^\top z^{\,i}, \quad g_i \in  \mathbb R^n,
\end{align}
which expresses how far $\mathcal {Z}_i$ can grow in the direction $g_i$ as measured by the projection $g_i^\top z^{\,i}$.
\hfill $\square$
\end{definition}

\begin{remark} \label{remark:RelTube}
    For the relative tubes $\mathcal Z_{ij}=\mathcal Z_i\oplus(-\mathcal Z_j)$ in Definition~\ref{def:tightened_safety_fxn}, it follows from Definition~\ref{def:support_fxn} that
\(\sigma_{\mathcal Z_{ij}}([g_i;g_j])=\sigma_{\mathcal Z_i}(g_i)+\sigma_{\mathcal Z_j}(g_j)\).
\end{remark}

\begin{lemma}\label{lem:tightened_safety_fxn}
Under the Assumption~\ref{assum:ConvexSafetyFunctions} and the assumptions of Lemmas~\ref{lem:RPI_tube},
\begin{align}\label{eq:h_tightened}
h_{ij}(x^{\,i},x^{\,j}) &\ge h_{ij}(\bar x^{\,i},\bar x^{\,j}) - \sigma_{\mathcal {Z}_i} (g_i) - \sigma_{\mathcal {Z}_j} (g_j),
\\
\label{eq:h_tightened_obs}
    h_{iO}(x^{i}) &\ge h_{iO}(\bar{x}^{i}) - \sigma_{\mathcal {Z}_i}\big(g_{iO}\big).
\end{align}
with $g_i := \nabla_{\bar x^i}h_{ij}(\bar x^i,\bar x^j)$, $g_j := \nabla_{\bar x^j}h_{ij}(\bar x^i,\bar x^j)$ and \mbox{$g_{iO} := \nabla_{\bar x^i}h_{iO}(\bar x^i)$}.
\hfill $\square$
\end{lemma}

\begin{pf}
Since the sets $(\mathcal Z_{i},\mathcal Z_{j})$ are closed, convex, and origin-symmetric, we have the dualities \citep[Section 2]{Hiriart2001}
\begin{equation} \label{eq:InfSupDuality}
    \inf_{z^i\in \mathcal {Z}_i} g_i^\top z^i = -\sup_{z^i\in\mathcal{Z}_i} (-g_i)^\top z^i \hspace{-3pt}\overset{\begin{array}{c}
\text{\tiny Def.}\\[-5pt]
\text{\tiny \ref{def:support_fxn}}
\end{array}}{=}\hspace{-3pt} -\sigma_{\mathcal{Z}_i}(-g_i) \hspace{-6pt}\overset{\begin{array}{c}
\text{\tiny origin}\\[-5pt]
\text{\tiny sym.}
\end{array}}{=}\hspace{-6pt} -\sigma_{\mathcal {Z}_i}(g_i)
\end{equation}

By the supporting-hyperplane property for convex $h_{ij}$ \citep[Equation 3.2]{boyd2004convex}, we can write the first-order global lower bound as
\begin{multline}
\hspace{-9pt}h_{ij}(\bar{x}^{i}+z^{i}, \bar{x}^{j}+z^{j}) \geq h_{ij}(\bar{x}^{i}, \bar{x}^{j}) + \nabla h_{ij}(\bar{x}^{i}, \bar{x}^{j})^\top \begin{bmatrix}z^{i}, z^{j}\end{bmatrix}^\top 
\\ 
\ge h_{ij}(\bar{x}^{i}, \bar{x}^{j}) + \nabla_{\bar x^i} h_{ij}(\bar{x}^{i}, \bar{x}^{j})^\top z^{i} + \nabla_{\bar x^j} h_{ij}(\bar{x}^{i}, \bar{x}^{j})^\top z^{j} \hfill
\\
\ge h_{ij}(\bar{x}^{i}, \bar{x}^{j}) + \inf_{z^i}\nabla_{\bar x^i} h_{ij}(\bar{x}^{i}, \bar{x}^{j})^\top z^{i} + \inf_{z^j} \nabla_{\bar x^j} h_{ij}(\bar{x}^{i}, \bar{x}^{j})^\top z^{j}
\\
\underset{\eqref{eq:InfSupDuality}}{\ge }
h_{ij}(\bar x^{\,i},\bar x^{\,j}) - \sigma_{\mathcal {Z}_i}\big(g_i\big) -  \sigma_{\mathcal {Z}_j}\big(g_j\big).
\end{multline}
for $g_i := \nabla_{\bar x^i}h_{ij}(\bar x^i,\bar x^j), g_j := \nabla_{\bar x^j}h_{ij}(\bar x^i,\bar x^j)$.

Similar arguments are used to obtain \eqref{eq:h_tightened_obs} for $h_{iO}(x^{i})$.
\hfill $\blacksquare$
\end{pf}

\begin{lemma}[Support Function - Tubes]\label{lem:support_fxn}
The support function of the RPI tube $\mathcal Z_i := \{\,z^{\,i}:~ (z^{\,i})^\top P_i z^{\,i}\le \rho_i^{\,2}\,\}$, with $\rho_i$ and $P_i\succ0$ satisfying the Lyapunov equation \eqref{eq:Lyapunov_eq_1} as in Lemma~\ref{lem:RPI_tube}, is
\begin{align}\label{eq:support_fxn}
\sigma_{\mathcal{Z}_i}(g_i) \;=\; \rho_i\, \sqrt{\,g_i^\top P_i^{-1} g_i}.
\end{align}
\hfill $\square$
\end{lemma}

\begin{pf}
Applying the definition~\ref{def:support_fxn} to the  tube $\mathcal Z_i$, we have
\[
\sigma_{\mathcal{Z}_i}(g_i) = \max_{z^i}\; g_i^\top z^i
\quad \text{s.t.} \quad
(z^i)^\top P_i z^i \le \rho_i^{2}
\]
Defining the Lagrangian
\[
\mathcal{L}(z^i,\lambda) \;=\; g_i^\top z^i \;-\; \lambda \bigl( (z^i)^\top P_i z^i - \rho_i^{2} \bigr),
\qquad \lambda \ge 0.
\]
and seeking its critical points, we obtain
\begin{align*}
\hspace{-35pt} \frac{\partial \mathcal{L}}{\partial z^i} = g_i - 2\lambda P_i z^i = 0
\quad \Rightarrow \quad
(z^i)^\star = \frac{1}{2\lambda} P_i^{-1} g_i,
\end{align*}
\begin{align*}
\hspace{45pt} \frac{\partial \mathcal{L}}{\partial \lambda} = 0
\quad \Rightarrow \quad ((z^i)^\star)^\top P_i (z^i)^\star = \rho_i^{2}.
\end{align*}
Therefore,
\[
\left(\frac{1}{2\lambda} P_i^{-1} g_i\right)^\top
P_i
\left(\frac{1}{2\lambda} P_i^{-1} g_i\right)
\;=\;
\frac{1}{4\lambda^{2}}\, g_i^\top P_i^{-1} g_i
\;=\;
\rho_i^{2}.
\]
which establishes that 
\(
\lambda \;=\; \dfrac{1}{2\rho_i}\, \sqrt{\,g_i^\top P_i^{-1} g_i\,}
\)
as~well~as $(z^{i})^{\star}=\dfrac{\rho_{i}}{\hspace{-4pt}\sqrt{g_{i}^{\top}P_{i}^{-1}g_{i}}}P_{i}^{-1}g_{i}$, thus giving
the support function
\begin{align*}
\sigma_{\mathcal{Z}_i}(g_i)
=
g_i^\top (z^i)^\star
=
g_i^\top \!\left(\frac{1}{2\lambda} P_i^{-1} g_i\right)
=
\rho_i \sqrt{g_i^T P_i^{-1} g_i}.
\end{align*}
\hfill $\blacksquare$
\end{pf}

\begin{figure}[htbp]
\centering
\begin{tikzpicture}[scale=0.95]

\definecolor{agenti}{RGB}{30,64,175}
\definecolor{agentj}{RGB}{6,95,70}
\definecolor{tubei}{RGB}{59,130,246}
\definecolor{tubej}{RGB}{16,185,129}
\definecolor{safety}{RGB}{190,30,90}
\definecolor{tightened}{RGB}{245,158,11}

\coordinate (center) at (0,0);          
\def\safetyRadius{1.2}

\coordinate (xi) at (-2.48,1.65);
\coordinate (xj) at ( 2.80,-1.5);
\coordinate (xjb) at ( 4.13,-0.5);

\def\tubeRadiusi{1.35}
\def\tubeRadiusj{0.8}

\pgfmathsetmacro{\sigmai}{\tubeRadiusi}
\pgfmathsetmacro{\sigmaj}{\tubeRadiusj}
\pgfmathsetmacro{\totalmargin}{\sigmai + \sigmaj}
\pgfmathsetmacro{\tightenedRadius}{\safetyRadius + \totalmargin}


\draw[rotate=50, dashed, thick] (0,0) ellipse (2cm and 0.9cm);
\draw[tightened, rotate=50, dashed, thick] (0,0) ellipse (3.2cm and 2.4cm);
\filldraw[black] ($(xi)+(-0.80,-0.55)$) circle (2pt);

\node[black, font=\tiny] at (0.9,1.9) {$h_{iO} (\bar x^{\,i})=0$};

\node[tightened, font=\tiny] at (1.25,3.15) {$h_{iO}^{\text{tight}}(\bar x^{\,i}) = 0$};

\filldraw[tubei, fill opacity=0.15, thick] (xi) circle (\tubeRadiusi);
\node[tubei, font=\small] at ($(xi)+(0.0,1.55)$) {$Z_i$};

\node[tubej, font=\tiny] at ($(xj)+(-2.8,1.5)$) {$\textbf{\text{Obstacle}}$};
\filldraw[tubej, fill opacity=0.15, thick]
    ($(0,0)+(1.3,1.4)$) --
    ($(0,0)+(-0.6,0.6)$) --
    ($(0,0)+(-1.3,-1.4)$) --
    ($(0,0)+(0.6,-0.6)$) -- cycle;

\filldraw[agenti] (xi) circle (2.5pt);
\node[agenti, font=\tiny] at ($(xi)+(0.3,0.25)$) {$\bar{x}_1^i(t)$};

\filldraw[agenti] (xi) circle (2.5pt);
\node[black, font=\tiny] at ($(xi)+(-0.50,-0.85)$) {${x}_1^i(t)$};



\draw[->, >=Stealth, black!70, line width=0.7pt,
      shorten <=34.0pt, shorten >=15.0pt]
  ($(xi)+(1.00,0.75)$) -- ($(xi)+(-1.25,-0.86)$)
  node[pos=0.55, inner sep=1pt, font=\small,  xshift=-6pt, yshift=1pt] {$z_1^i$};












\draw[-{Stealth[length=3mm]}, tubei, line width=1.2pt]
    (xi) -- ++(1.70,-0.9)
    node[midway, above, xshift=27.5pt, yshift=-7pt, font=\tiny] {$-\nabla_{\bar{x}^i} h_{iO}$};


\draw[safety, dashed, line width=1.8pt] (xi) -- ++(1.20,-0.66);
\filldraw[safety] ($(xi)+(1.2,-0.66)$) circle (1.8pt);
\node[safety, font=\scriptsize\bfseries] at ($(xi)+(1.30,-1.00)$) {$\sigma_{Z_i}$};







\begin{scope}[shift={(2.0,-2.89)}]
[shift={(-2.2,-0.5)}]
\draw[black!40, line width=0.8pt, rounded corners]
    (0.0,-0.17) rectangle (3.30,1.6);


\filldraw[agenti] (0.4,1.05) circle (2pt);
\node[font=\tiny\tiny, anchor=west] at (0.45,1.05) {Nominal trajectory};

\filldraw[black]  (0.4,1.35) circle (2pt);
\node[font=\tiny\tiny, anchor=west] at (0.45,1.28) {Actual trajectory};

\draw[tubei, thick] (0.4,0.80) circle (0.1);
\node[font=\tiny\tiny, anchor=west] at (0.45,0.80) {RPI tube $Z_i$};

\draw[tubej, thick] (0.4,0.55) circle (0.1);
\node[font=\tiny\tiny, anchor=west] at (0.45,0.55) {Obstacle};

\draw[black, dashed, thick] (0.3,0.30) -- (0.7,0.30);
\node[font=\tiny\tiny, anchor=west] at (0.6,0.30) {Safety boundary};

\draw[tightened, dashed, thick] (0.3,0.10) -- (0.7,0.10);
\node[font=\tiny\tiny, anchor=west] at (0.6,0.05) {Tightened boundary};
\end{scope}

\draw[->, black!50, opacity=0.4, line width=0.8pt] (-4.0,0) -- (4.5,0) node[below right] {$x_{1,1}$};
\draw[->, black!50, opacity=0.4, line width=0.8pt] (0,-3.0) -- (0,3.0) node[above left] {$x_{1,2}$};

\end{tikzpicture}

\caption{Geometric illustration of tube-based tightening for agent--obstacle avoidance in 2D. The nominal state $\bar x^i(t)$ is surrounded by an RPI tube $Z_i$ and $x^i(t)$. The control barrier function boundary $h_{iO}$ (black dashed) is tightened by the worst-case decrease of $h_{iO}$ under tube deviations, bounded by $\delta_{iO}(\bar x^i)=\sigma_{Z_i}(\nabla_{\bar x^i}h_{iO})$. Enforcing $h_{iO}(\bar x^i)\ge \delta_{iO}(\bar x^i)$ (equivalently $h_{iO}-\delta_{iO}\ge 0$, orange dashed) guarantees $h_{iO}(x^i)\ge 0$ for all admissible tube realizations.} \label{fig:tube_tightening}
\end{figure}


\section{Main Result}\label{sec:MainResult}
Building upon the methodology presented in Section~\ref{Methodology}, we now detail the tube-MPC with tightened eCBF framework that performs optimization on the nominal trajectories.
We enforce the tightened eCBF inequalities by replacing $h_{ij}^{\text{tight}}(\cdot)$ and $h_{iO}^{\text{tight}}(\cdot)$ for the zero-order term $\kappa_0 h_{ij}(\cdot)$, while keeping all higher-order Lie derivatives evaluated on $h_{ij}(\cdot).$

Following the standard eCBF methodology \citep{Nguyen2016}, we construct exponential control barrier functions $\Phi_{ij}$ and $\Phi_{iO}$ from $h_{ij}(\bar{x}^{i},\bar{x}^{j})$ and $h_{iO}(\bar{x}^{i})$, 
\begin{align}\label{eq:std_ecbf_ij}
\Phi_{ij}(x^i, x^j,u^i) &:= L_{F_{x}}^{r}h_{ij}(x^i, x^j) + \sum_{q=0}^{r-1}\kappa_q L_{F_{x}}^{q}h_{ij}(x^i, x^j) \notag \\&  + L_{F_{u}}L_{F_{x}}^{r-1}h_{ij}(x^i, x^j) u^i ,
\\
\label{eq:std_ecbf_iO}
\Phi_{iO}(x^i,u^i) &:= L_{F_{x}}^{r}h_{iO}(x^i) + \sum_{q=0}^{r-1}\kappa_q L_{F_{x}}^{q}h_{iO}(x^i) \notag \\& + L_{F_{u}}L_{F_{x}}^{r-1}h_{iO}(x^i) u^i .
\end{align}

\subsection{Proposed Tightened eCBF Functions}

We propose the following tightened exponential control barrier functions (tight-eCBF) using the standard eCBF 
\eqref{eq:std_ecbf_ij}–\eqref{eq:std_ecbf_iO}, to impose on the nominal trajectories of agents while ensuring safety of the actual trajectories which are subject to unknown disturbances $w^i$ and corresponding errors $z^i$ as
\begin{multline}\label{eq:eCBF_col_final}
\hspace{-9pt} \Phi_{ij}^{\text{tight}}\big(\bar{x}^{i}, \bar{x}^{j}, \bar{u}^{i}\big)=L_{F_{x}}^{r} h_{ij}(\bar{x}^{i},\bar{x}^{j}) + \sum_{q=1}^{r-1} \kappa_q L_{F_{x}}^{q} h_{ij}(\bar{x}^{i},\bar{x}^{j}) \\  \hspace{9pt} + \big(L_{F_{u}} L_{F_{x}}^{r-1} h_{ij}(\bar{x}^{i},\bar{x}^{j})\big) \bar{u}^{i}
+\kappa_0 \big( h_{ij} (\bar x^i,\bar x^j) -\delta_{ij}(\bar{x}^{i},\bar{x}^{j})\big),
\end{multline}
\begin{multline}\label{eq:eCBF_obs_final}
\Phi_{iO}^{\text{tight}}\big(\bar{x}^{i}, \bar{u}^{i}\big) = L_{F_{x}}^r h_{iO}(\bar{x}^{i}) + \sum_{q=1}^{r-1} \kappa_q L_{F_{x}}^{q} h_{iO}(\bar{x}^{i}) 
\\
+ \big(L_{F_{u}} L_{F_{x}}^{r-1} h_{iO}(\bar{x}^{i})\big) \bar{u}^{i} + \kappa_0 \big( h_{iO} (\bar x^i) - \delta_{iO}(\bar{x}^{i})\big),
\end{multline}
where $\kappa_0,\dots,\kappa_{r-1} > 0$ are eCBF gains for a relative-degree-$r$ safety function, and where we use notations $\delta_{ij}(\bar{x}^{i},\bar{x}^{j}):=\sigma_{\mathcal{Z}_{i}}(g_{i})+\sigma_{\mathcal{Z}_{j}}(g_{j})\equiv\rho_i\left\Vert \nabla_{\bar{x}^{i}}h_{ij}(\bar{x}^{i},\bar{x}^{j})\right\Vert _{P_{i}^{-1}}+\rho_j\left\Vert \nabla_{\bar{x}^{j}}h_{ij}(\bar{x}^{i},\bar{x}^{j})\right\Vert _{P_{j}^{-1}}$, together with 
\noindent\(\delta_{iO}(\bar{x}^{i}):=\sigma_{\mathcal Z_i}(g_{iO}) \equiv {\rho_i}\left\Vert \nabla_{\bar{x}^{i}}h_{iO}(\bar{x}^{i},\bar{x}^{j})\right\Vert _{P_{i}^{-1}}\).

The following theorem establishes conditions for forward invariance of the true states set $\mathcal{C}$. 
\begin{theorem} [Forward Invariance of Set] \label{thm:tube_mpc_ecbf}
Let $r \ge 1$ be the relative degree of the eCBF $h_{ij}(\cdot,\cdot)$ and $h_{iO}(\cdot)$ as in Definition~\ref{def:RelDeg}. Let the nominal input processes $\left(\bar{u}^{1},\cdots,\bar{u}^{N}\right)$ and the corresponding nominal trajectories $\left(\bar{x}^{1},\cdots,\bar{x}^{N}\right)$, satisfy the inequalities $\Phi_{ij}^{\text{tight}}\big(\bar{x}^{i}, \bar{x}^{j}, \bar{u}^{i}\big) \geq 0$ and $\Phi_{iO}^{\text{tight}}\big(\bar{x}^{i}, \bar{u}^{i}\big) \geq 0$ for the tightened safety functions \eqref{eq:eCBF_col_final} and \eqref{eq:eCBF_obs_final}, for some positive gains $\kappa_0,\ldots,\kappa_{r-1}$ so the polynomial $p(s)=s^r+\kappa_{r-1}s^{r-1}+\cdots+\kappa_1 s+\kappa_0 \equiv \prod_{q=1}^{r}(s+c_q)$ is Hurwitz (i.e., $c_q > 0$ for all $q$). 

Suppose that at the initial time $t_0$ the safety constraints 
$h_{ij}(x^{i}(t_0), x^{j}(t_0)) \geq 0$, $h_{iO}(x^{\,i}(t_0)) \ge 0$, together with $L_{F_x}^k h_{ij}(x^{i}(t_0), x^{j}(t_0)) \geq 0$, $L_{F_x}^k h_{iO}(x^{i}(t_0), x^{j}(t_0)) \geq 0$, $k \in \{1, \cdots, r\}$ are satisfied, and the initial error is within the safety tube, i.e., $z^i(t_0) \in \mathcal{Z}_i$. Then, under the Assumptions of Lemmas~\ref{lem:RPI_tube}--\ref{lem:support_fxn},
\begin{align} \label{eq:SafetyGaurantee}
    &h_{ij}(x^{\,i}(t),x^{\,j}(t)) \ge 0, \quad h_{iO}(x^{\,i}(t)) \ge 0,
\end{align}
for all $t\ge t_0$, i.e., the collision-avoidance and obstacle-avoidance set \eqref{eq:SafetySet} for the state of all agents are forward invariant.
\hfill $\square$
\end{theorem}

\begin{pf}
By Lemma \ref{lem:tightened_safety_fxn}, we obtain from  \eqref{eq:h_tightened} that
\[
h_{ij}(x^i,x^j)\ \ge\ h_{ij}(\bar x^i,\bar x^j)-\delta_{ij}(\bar x^i,\bar x^j).
\]
where the support function \eqref{eq:support_fxn}, provided in Lemma \ref{lem:support_fxn}, are bounded by 
\[ 
\sigma_{\mathcal Z_i}(g)=\rho_i\sqrt{g^\top P_i^{-1}g}\ \le\ \rho_i\|P_i^{-1/2}\|\|g\|.
\] 
for $g := \nabla_{\bar x^i}h_{ij}(\bar x^i,\bar x^j)$.
Therefore, for all \(t\ge t_0\), 
\begin{multline}
\delta_{ij}(\bar x^i(t),\bar x^j(t))
=\sigma_{\mathcal Z_i}(\nabla_{\bar x^i}h_{ij})+\sigma_{\mathcal Z_j}(\nabla_{\bar x^j}h_{ij}) 
\\ 
\le\ \underbrace{\rho_i\|P_i^{-1/2}\|G_i^{\max}+\rho_j\|P_j^{-1/2}\|G_j^{\max}}_{:=\ \Delta_{ij}}. 
\end{multline} 
where $G_i^{\max}$ and $G_j^{\max}$ are some finite bounds on the gradients of $h_{ij}$ with respect to $\bar x^i$ and $\bar x^i$, i.e.,
\[ 
\|\nabla_{\bar x^i} h_{ij}(\bar{x}^{i},\bar{x}^{j})\|\le G_i^{\max},\qquad
\|\nabla_{\bar x^j} h_{ij}(\bar{x}^{i},\bar{x}^{j})\|\le G_j^{\max}
\] 
which exist because $h_{ij}$ is continuously differentiable under Assumption~\ref{assum:ConvexSafetyFunctions}, and $\bar x^i$ and $\bar x^i$ are bounded reference trajectories.
Hence
\[ 
\ \delta_{ij}(\bar x^i,\bar x^j)\ \le\ \Delta_{ij}\quad \forall t\ge t_0. 
\] 
Accordingly, we obtain the bound on the tight-eCBF  \eqref{eq:eCBF_col_final} as
\begin{multline}\label{eq:tight-ecbf}
\Phi_{ij}^{\text{tight}}\big(\bar{x}^{i}, \bar{x}^{j}, \bar{u}^{i}\big)=L_{F_{x}}^{r} h_{ij}(\bar{x}^{i},\bar{x}^{j}) + \sum_{q=1}^{r-1} \kappa_q L_{F_{x}}^{q} h_{ij}(\bar{x}^{i},\bar{x}^{j}) \\  \hspace{9pt} + \big(L_{F_{u}} L_{F_{x}}^{r-1} h_{ij}(\bar{x}^{i},\bar{x}^{j})\big) \bar{u}^{i}
+\kappa_0 \big( h_{ij} (\bar x^i,\bar x^j) -\delta_{ij}(\bar{x}^{i},\bar{x}^{j})\big)
\\
\geq  L_{F_{x}}^{r} h_{ij}(\bar{x}^{i},\bar{x}^{j}) + \sum_{q=1}^{r-1} \kappa_q L_{F_{x}}^{q} h_{ij}(\bar{x}^{i},\bar{x}^{j}) \\  \hspace{-10pt} + \big(L_{F_{u}} L_{F_{x}}^{r-1} h_{ij}(\bar{x}^{i},\bar{x}^{j})\big) \bar{u}^{i}
+ \kappa_0 \big(h_{ij}(\bar{x}^{i},\bar{x}^{j})- \Delta_{ij}\big) . 
\end{multline}

Therefore, defining the auxiliary eCBF as
\(
\tilde{h}_{ij} := h_{ij}  - \Delta_{ij},
\)
and invoking \citep[Theorem 2]{Nguyen2016} 
we deduce that $\tilde{h}_{ij}^{(k)}(\bar x^i(t),\bar x^j(t)) \geq 0$ for all $t \geq t_0$, which subsequently establishes that the positivity of $\Phi_{ij}^{\text{tight}}$ in \eqref{eq:tight-ecbf} yields
\begin{align}\label{eq:tilde_h_cbf}
h_{ij}(\bar x^i,\bar x^j) \geq \Delta_{ij}, \; \forall t\ge t_0.
\end{align}
Substituting \eqref{eq:tilde_h_cbf} into \eqref{eq:h_tightened} from Lemma~\ref{lem:tightened_safety_fxn}, gives
\begin{align}
    h_{ij}( x^i(t), x^j(t))\ &\ge\ h_{ij}(\bar x^i,\bar x^j) -\delta_{ij}(\bar x^i,\bar x^j) \notag \\ &
    \ge\Delta_{ij} - \delta_{{ij}}(\bar x^i,\bar x^j) \ge 0, \quad \forall t \ge t_0
\end{align}
Repeating similar steps for the obstacle safety functions $h_{iO}$, establishes that $h_{iO}( x^i)\ge 0$ for all $O \in \mathcal{O}_i$. In other words, the true local safety set $\mathcal{C}$ defined in \eqref{eq:SafetySet} is forward invariant, i.e.,
\[x^i(t) \in \mathcal{C}\]
for all $t\ge t_0$.
\hfill $\blacksquare$
\end{pf}

\subsection{Distributed Tube-MPC eCBF Formation Framework}

Having established that the enforcement of the proposed tightened safety functions \eqref{eq:eCBF_col_final} and \eqref{eq:eCBF_obs_final} on nominal trajectories of the agents ensures that the actual trajectories of the system are safe, we now present a Distributed Tube-MPC with eCBF which generates the nominal trajectories.

To avoid requiring all-to-all communication, reduce computational load at agent level, and enable distributed control policies, we consider a communication graph \mbox{$\mathcal{G}=(\mathcal{V},\mathcal{E},A)$} with the node set
$\mathcal{V}=\{1,\dots,N\}$, edge~set $\mathcal{E}\subset\mathcal{V}\times\mathcal{V}$, and
weighted adjacency matrix $A=[a_{ij}]\in\mathbb{R}^{N\times N}$, where
\[
  a_{ij} :=
  \begin{cases}
    w_{ij} & \text{if } (j,i)\in\mathcal{E},\\
    0 & \text{otherwise},
  \end{cases}
\]
where $w_{ij}>0$ for all $(j,i)\in\mathcal{E}$. For the special case of undirected graphs, $A$ is symmetric and $w_{ij}=w_{ji}$.
We define the neighboring agents set $\mathcal{N}_i:=\{\,j\in\mathcal{V}: a_{ij}>0\,\}$.


To enable decentralized communication while avoiding unsafe loss of information in safety-critical situations, we impose the following information architecture.

\begin{assumption}\label{ass:prox_threshold}
    The communication weights has the property that, if 
    \(\Vert \bar{x}^i_1 - \bar{x}^j_1 \Vert \leq \phi\) 
    for some position proximity threshold \(\phi \in \mathbb{R}_{>0}\), then \(a_{ij} \neq 0\). \hfill $\square$
\end{assumption}

The threshold \(\phi\), specified as part of the problem setting, is chosen in practice to be no smaller than the safety distance plus the tube-induced safety margins, ensuring that any agents that may approach the safety boundary can exchange the information required for safety enforcement.


The in-degree matrix is denoted by $D=\mathrm{diag}\{d_i\}$ and defined via $d_i=\sum_{j=1}^N a_{ij}$, and the Laplacian~is~\mbox{$L:=D-A$}. 

The leader’s interaction matrix is defined as $B_0=\mathrm{diag}\{b_{i0}\}\in\mathbb{R}^{N\times N}$ with
\[
  b_{i0} :=
  \begin{cases}
    w_{i0} & \text{if } (0,i)\in\bar{\mathcal{E}},\\
    0 & \text{otherwise},
  \end{cases}
  \qquad w_{i0}>0.
\]
We introduce a leader node~$0$ and the augmented graph $\bar{\mathcal{G}}=(\bar{\mathcal{V}},\bar{\mathcal{E}})$ with $\bar{\mathcal{V}}=\{0,1,\dots,N\}$.
The augmented Laplacian used in leader-follower formation control is defined as $L_{B_0} := L + B_0$. In order to ensure that no clusters of agents are isolated from the leader, we impose the following assumption. 
\begin{assumption} \label{asm:SpanningTree}
The augmented graph \( \bar{\mathcal{G}} \) contains a spanning tree with the leader as the root, i.e., for every agent~$i$ there exists a (directed) path in $\bar{\mathcal{G}}$ from the leader node $v_0$ to $v_i$. In other words, whenever $(v_i, v_0) \notin \bar{\mathcal{E}}$, there exists a sequence of nonzero elements of \(A\) of the form ${a}_{i i_2}, {a}_{i_2 i_3}, \cdots, {a}_{i_{l-1} i^{\prime}}$, for some $(i^{\prime},0) \in \bar{\mathcal{E}}$, with the sequence length $l$ being a finite integer.
\hfill $\square$
\end{assumption}


To accommodate tracking with offsets and ensure coordinated movement while maintaining formation, we denote by \(\psi_p^{i} \in \mathbb{R}^{d}\) and \(\psi_p^{0} \in \mathbb{R}^{d} \) desired state offset (formation geometry) from the leader or other agents. 
As presented in \citep{koulong2025acc} for the distribution of policies, 
we define the local weighted stability error of each agent $i$ as
 \begin{align}
     r^i &= - \nu_1 \sum_{p=1}^{n} \sum_{j=1}^{N} \lambda_{p} a_{ij} \bigg[(x_{p}^{i} - \psi_{p}^{i}) - (x_{p}^{j} - \psi_{p}^{j})\bigg] \notag \\
&- \nu_2 \sum_{p=1}^{n} \lambda_{p}  b_{i0} \bigg[(x_{p}^{i} - \psi_{p}^{i}) - (x_{p}^{0} - \psi_{p}^{0})\bigg],
 \end{align}
with arbitrary positive constants $\nu_1, \nu_2 >0$, and design parameters $\lambda_j,\; j = 1, \ldots, n-1$ with the restriction that the associated characteristic polynomial $s^{n-1} + \lambda_{n-1} s^{n-2} + \cdots + \lambda_1$ is Hurwitz.  

Using Theorem~\ref{thm:tube_mpc_ecbf}, we pose a DMPC problem with tightened eCBF constraints, solved via continuous-time transcription (multiple shooting with RK4). At each sampling time \(t_k\), we optimize the nominal sequence \((\bar x^{i},\bar u^{i})\) over horizon \(H\) with step \(T_s\), enforce the tightened eCBF inequalities on \(\bar x^{i}\), and apply the tube feedback to the plant. For a given \(H\in\mathbb{N}\), we minimize the objective subject to \eqref{eq:ct_follower_nominal} for \(k=0,\ldots,H-1\): 
\begin{align}
\min_{\{\bar{u}^{i}_{k}\}_{k=0}^{H-1}}~&
\sum_{k=0}^{H-1}
\Big(
\|\bar{r}^{i}_{k}\|_{Q_r}^2
+\|\bar{u}^{i}_{k}\|_{R}^2
+\|\Delta \bar{u}^{i}_{k}\|_{R_\Delta}^2
\Big) 
+ \|\bar{r}^{i}_{H}\|_{P_r}^2 \label{eq:cost} \\[1mm]
\text{s.t. }~~& \notag \\
&\bar{x}^{i}_{k+1} = \Phi_{\text{RK4}}\big(\bar{x}^{i}_{k}, \bar{u}^{i}_{k}, T_s\big),  \label{eq:dynamics_constraint}\\
&\bar{x}^{i}_{0} = \bar{x}^{i}(t), \label{eq:initial_condition}\\
&\big(\bar{x}^{i}_{k}, \bar{u}^{i}_{k}\big) \in \big(\mathcal{X}_i \ominus \mathcal{Z}_i\big) \times \big({U}_i \ominus K^{i}\mathcal{Z}_i\big),  \label{eq:state_input_constraints}\\
&\Phi_{ij}^{\text{tight}}\big(\bar{x}^{i}_{k}, \bar{x}^{j}_{k}, \bar{u}^{i}_{k}\big) \ge 0, \; \forall (i,j) \in \mathcal{E}, \label{eq:interagent_tight}\\
&\Phi_{iO}^{\text{tight}}\big(\bar{x}^{i}_{k}, \bar{u}^{i}_{k}\big) \ge 0, \; \forall (i,O) \in \mathcal{O}^{\text{act}}, \label{eq:obstacle_tight}\\
&\bar{x}^{i}_{H} \in \bar{\mathcal{X}}_{f,i}, \label{eq:terminal_constraint}
\end{align}
where $\Delta \bar{u}^{i}_{k} := \bar{u}^{i}_{k} - \bar{u}^{i}_{k-1}$, $(Q_r, P_r, R, R_\Delta) \succ 0$ are weighting matrices, and $\Phi_{ij}^{\text{tight}}$, $\Phi_{iO}^{\text{tight}}$ are defined in \eqref{eq:eCBF_col_final} and \eqref{eq:eCBF_obs_final}, respectively.
The terminal safe set is defined as
\begin{multline}
\bar{\mathcal{X}}_{f,i} 
:= \Bigg\{ \bar{x}^{i}_f \in \bar{\mathcal{X}}_i \,\text{ s.t. } 
\\
\Phi^{\text{tight}}_{iO}\big(\bar{x}^{i}_f, \bar{u}^{i}_f\big) \ge 0, \; \forall (i,O) \in \mathcal{O}^{\text{act}}
\\
\Phi^{\text{tight}}_{ij}\big(\bar{x}^{i}_f, \bar{x}^{j}_f, \bar{u}^{i}_f\big) \ge 0, \; \forall j \in \mathcal{N}_i, 
\exists \bar{u}^{i}_f \in \bar{U}_i 
\Bigg\}
\end{multline}

The satisfaction of the constraints within the MPC optimization—particularly the tightened eCBF conditions \eqref{eq:interagent_tight} and \eqref{eq:obstacle_tight} and the terminal constraint \eqref{eq:terminal_constraint} —provides a recursively feasible nominal control inputs. By virtue of Theorem~\ref{thm:tube_mpc_ecbf}, implementing this policy using the composite control law \eqref{eq:ancillary_control} guarantees the forward invariance of the safe set for the true system dynamics, preventing collisions despite external disturbances.

\section{Numerical Example}\label{NUMERICALEXAMPLE} 

\begin{figure}[htbp]
    \centering
    \includegraphics[width=2.5in]
    {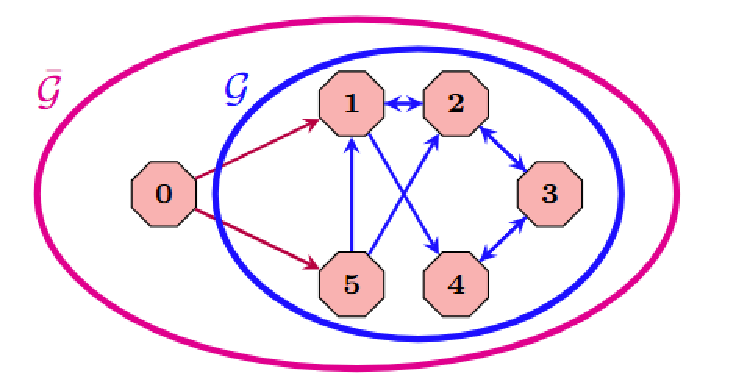}
    \caption{The considered communication topology \({\mathcal{G}}\) and the augmented graph \(\bar{\mathcal{G}}\) for the example in Section~\ref{NUMERICALEXAMPLE}.}
    \label{fig:1}
\end{figure}

We validate the framework on a 2D leader–follower formation with one leader and \mbox{$N=5$} followers, each modeled as a third-order $n=3$ nonlinear integrator-chain system with bounded disturbances. Simulations run for \( 30 \) s with \( T_s = 0.1 \) s \( (300 \) steps) and DMPC horizon \( H = 5 \). Two circular obstacles are placed at \( [1,1] \) and \( [-1.5,0.5] \) with radii \( 0.50 \) and \( 0.65 \) (plus \( 0.15 \) inflation). Ancillary gains are \( K_p = -[15,4,15,8,6,8] \) (pole placement), formation offsets are $\psi=[[-9;2], [-6;2], [0;2], [6;2], [9;2]]$, and eCBF gains are \( \kappa_0 = 30, \kappa_1 = 38, \kappa_2 = 3 \). The leader starts at \( [3,0,0,0,0,0] \) and followers at $[1.0, -0.5, \text{zeros}(1,4)]$,~$[-0.8, 0.4, \text{zeros}(1,4)]$, \\~$[0.6, 0.6, \text{zeros}(1,4)]$, $[-0.5,  -0.7, \text{zeros}(1,4)]$, \\and~$[0.7, 0.0, \text{zeros}(1,4)]$. MPC weights are \( Q_r = 50 I_{10}, P_r = 10 I_{10}, R = 0.01 I_2, R_{du} = 0.001 I_2 \), with coupling \( \nu_1 = \nu_2 = 1 \) and stability parameters \( \lambda_1 = 100, \lambda_2 = 50 \). Figure~\ref{fig:1} depicts the information-flow structure.

\begin{figure}[htbp]
    \centering
    \begin{subfigure}[b]{0.410\textwidth}
    \centering
    \includegraphics[width=2.9in]
    {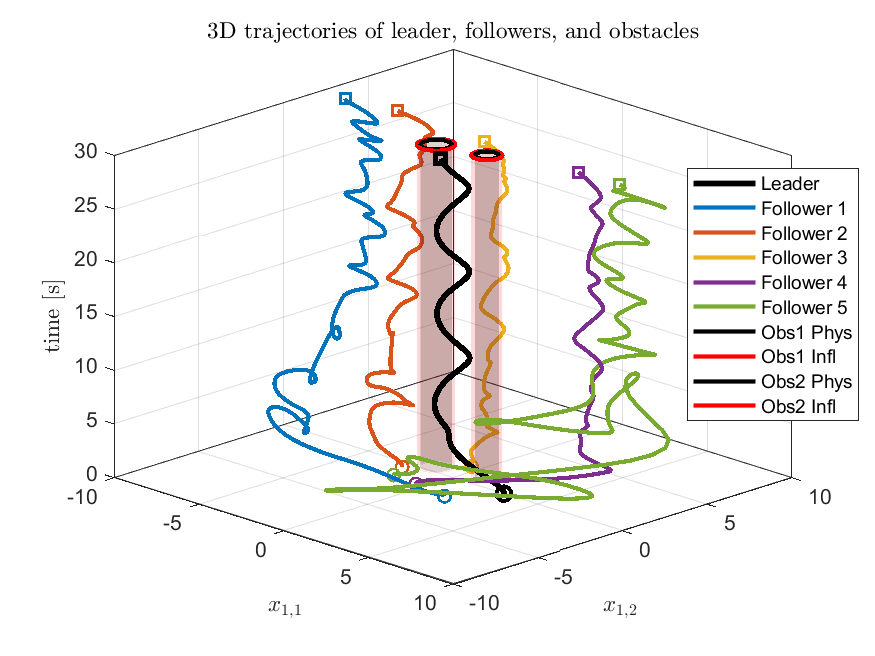}
    \caption{Leader and Follower Agents $[x^i_{1,1},x^i_{1,2},t]$}
    \label{fig:2}
    \end{subfigure}
    \hfill
    \begin{subfigure}[b]{0.410\textwidth}
    \centering
    \includegraphics[width=2.9in]{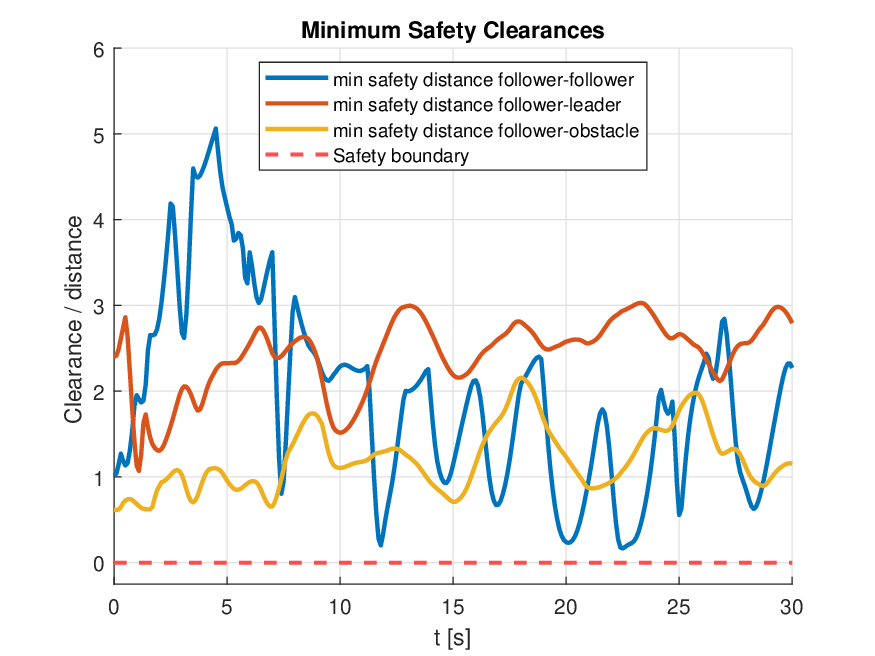}
    \caption{Minimum safety clearance}
    \label{fig:3}
    \end{subfigure}
    \caption{The corresponding evolution of the positional states $x^i_1 \in \mathbb{R}^2$, for the \mbox{leader $i=0$}, and follower agents  $i \in \{1,2,\cdots,5\}$ with respect to obstacles} 
    \label{fig:combined0}
\end{figure}

\begin{figure}[htbp]
    \centering
    \begin{subfigure}[b]{0.410\textwidth}
    \centering
    \includegraphics[width=2.9in]{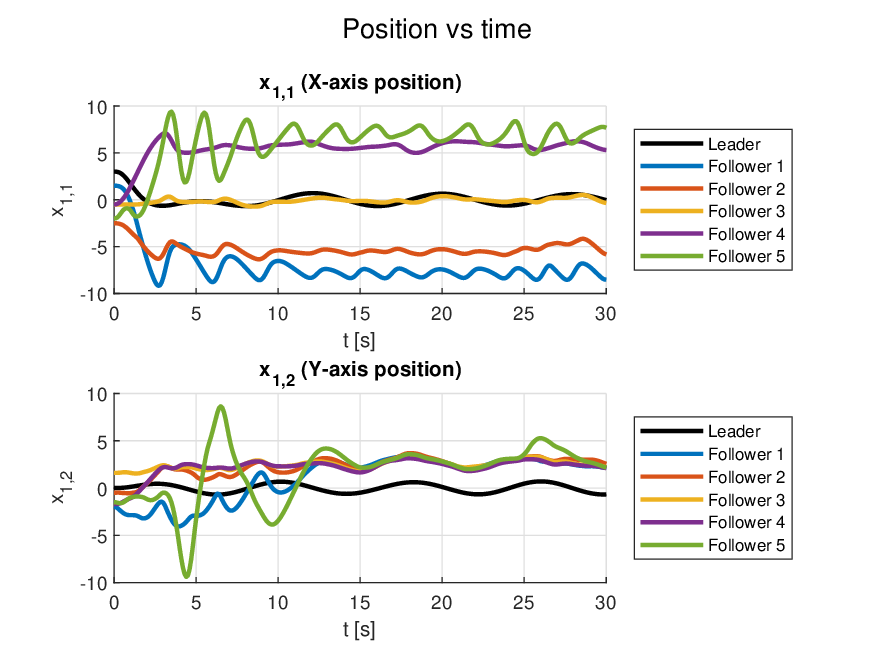}
    \caption{Leader and Follower Agents Component States $[x^i_{1,1},x^i_{1,2}]$}
    \label{fig:4}
    \end{subfigure}
    \hfill
    \begin{subfigure}[b]{0.410\textwidth}
    \centering
    \includegraphics[width=2.9in]{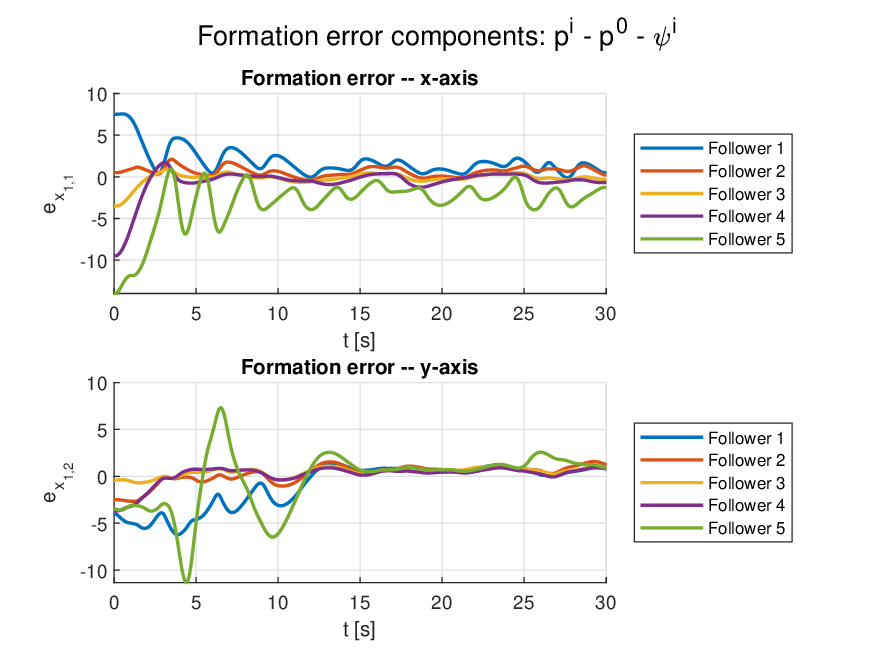}
    \caption{Leader and Follower Agents $[x_{1,1},x_{1,2}]$}
    \label{fig:5}
    \end{subfigure}
    \caption{Relative $[x^i_{1,1},x^i_{1,2}]$ state Error between Agents and Leader in y-direction}
    \label{fig:combined1}
\end{figure}

\scalebox{.71}{
\small
\setlength{\tabcolsep}{6pt}
\renewcommand{\arraystretch}{1.1} 
\hspace{-12pt}    \begin{tabular}{|c|c|}
\hline 
Agent  & $\dot{x}_{1,k}^{i} = x_{2,k}^{i}$,
$\dot{x}_{2,k}^{i} = x_{3,k}^{i}$, $k=1,2$, with $\dot{x}_{3,k}^{i} = f^i_k(x^i) + u^i_k + w^i_k$ presented below:  \tabularnewline
\hline 
\hline 
Leader & $\begin{aligned}
    f^0_1(x^0) &= -\,5\,\bigl(x_{3,1}^{0}\;+\;0.36\,x_{1,1}^{0}\;+\;0.84\,x_{2,1}^{0}\; +\; 0.15\,(x_{1,1}^{0})^3\; -\;0.4\,\sin\bigl(0.8\,t)\bigr),\\
f^0_2(x^0) &= -\,5\,\bigl(x_{3,2}^{0}\;+\;0.36\,x_{1,2}^{0}\;+\;0.84\,x_{2,2}^{0}\; \; 0.15\,(x_{1,2}^{0})^3\; -\;0.4\,\sin\bigl(0.8\,t + \frac{\pi}{2})\bigr),
\\
w_1^3(t) &= 0.20\,\sin(0.9\,t), \, w_2^3(t) = 0.25\,\sin\!\big(1.1\,t-\tfrac{\pi}{7}\big);\,\, u^0_1=u^0_2=0
\end{aligned}$ \tabularnewline
\hline 
Agent 1 & $\begin{aligned}
    f^1_1(x^1) = -k_a^1\!\big(x_{3,1}^1 + 0.49\,x_{1,1}^1 + 1.12\,x_{2,1}^1 + 0.12\,(x_{1,1}^1)^3   - 0.25\,\tanh(0.6\,x_{1,1}^1)\big), \\
    f^1_2(x^1) = -k_a^1\!\big(x_{3,2}^1 + 0.49\,x_{1,2}^1 + 1.12\,x_{2,2}^1 + 0.12\,(x_{1,2}^1)^3   - 0.25\,\tanh(0.6\,x_{1,2}^1)\big), \\
    w_1^1(t) = 0.20\,\sin(0.9\,t), \, w_2^1(t) = 0.15\,\sin\!\big(1.1\,t+\tfrac{\pi}{7}\big).

\end{aligned}$ \tabularnewline
\hline 
Agent 2 & $\begin{aligned}
    f^2_1(x^2) = -k_a^2\!\big(x_{3,1}^2 +0.36\,x_{1,1}^2 + 0.84\,x_{2,1}^2 + 0.18\,(x_{1,1}^2)^3  - 0.15\,(x_{1,2}^2)^2\,\tanh(x_{1,1}^2)\big), \\
f^2_2(x^2) = -k_a^2\!\big(x_{3,2}^2+0.36\,x_{1,2}^2 + 0.84\,x_{2,2}^2 + 0.18\,(x_{1,2}^2)^3   - 0.15\,(x_{1,1}^2)^2\,\tanh(x_{1,2}^2)\big), 
\\
w_1^2(t) = 0.18\,\sin(1.1\,t+0.3), \, w_2^2(t) = 0.18\,\sin\!\big(0.8\,t-\tfrac{\pi}{5}\big).
\end{aligned}$ \tabularnewline
\hline 
Agent 3 & $\begin{aligned}
    f^3_1(x^3) = -k_a^3\!\big(x_{3,1}^3+0.25\,\tanh(x_{1,1}^3) + 0.9\,\tanh(x_{2,1}^3) - 0.20\,\sin(0.7\,t)\big), \\
f^3_2(x^3) = -k_a^3\!\big(x_{3,2}^3+0.25\,\tanh(x_{1,2}^3) + 0.9\,\tanh(x_{2,2}^3)  - 0.20\,\sin(0.9\,t+\tfrac{\pi}{3})\big), 
\\
w_1^3(t) = 0.18\,\sin(1.1\,t+0.3), \, w_2^3(t) = 0.18\,\sin\!\big(0.8\,t-\tfrac{\pi}{5}\big).
\end{aligned}$ \tabularnewline
\hline 
Agent 4 & $\begin{aligned}
f^4_1(x^4) = -k_a^4\!\big(x_{3,1}^4+0.4225\,x_{1,1}^4 + 0.975\,x_{2,1}^4   + 0.08\,\big(x_{1,1}^4+x_{1,2}^4\big)^3\big), 
\\
f^4_2(x^4) = -k_a^4\!\big(x_{3,2}^4+0.4225\,x_{1,2}^4 + 0.975\,x_{2,2}^4 - 0.08\,\big(x_{1,1}^4-x_{1,2}^4\big)^3\big), \\
w_1^4(t) = 0.12\,\sin(0.6t), \, w_2^4(t) = 0.12 \sin\!\big(0.6t+\tfrac{\pi}{2}\big).
\end{aligned}$ \tabularnewline
\hline 
Agent 5 & $\begin{aligned}
f^5_1(x^5) = -k_a^5\!\big(x_{3,1}^5+0.3025\,x_{1,1}^5 + 0.88\,x_{2,1}^5 + 0.15\,(x_{1,1}^5)^3    - 0.20\,\tanh(0.5\,x_{1,1}^5)\big), \\
f^5_2(x^5) = -k_a^5\!\big(x_{3,2}^5+0.3025\,x_{1,2}^5 + 0.88\,x_{2,2}^5 + 0.15\,(x_{1,2}^5)^3    - 0.20\,\tanh(0.5\,x_{1,2}^5)\big), \\
w_1^5(t) = 0.22\,\sin\!\big(0.9\,t+\tfrac{\pi}{8}\big), \, w_2^5(t) = 0.20\,\sin(1.0\,t).
\end{aligned}$  \tabularnewline
\hline 
\end{tabular}
}

\subsection*{\textbf{Analysis}}
The simulation results demonstrate the core promise of the proposed framework: the synthesis of controllers that ensure safe, coordinated multi-agent motion in the presence of disturbances and obstacles.


Figure \ref{fig:combined0} shows that the leader and followers remain outside the obstacle regions while evolving along bounded trajectories over time. The 3D plot indicates that the followers maneuver around the obstacles rather than cutting through them, while still maintaining coherent motion relative to the leader.
Figure \ref{fig:3} confirms the safety result quantitatively: the minimum follower-follower, follower-leader, and follower-obstacle clearances all remain strictly above the zero safety boundary throughout the simulation. This indicates that no collisions or obstacle penetrations occur, and that the safety constraints remain effective over the full trajectory.


Figure \ref{fig:combined1} shows that the followers remain bounded and progressively align with the moving leader after the initial transient. The position plots indicate that the group recovers a coherent leader-relative formation, while the error plots show that the largest transients occur early and then contract to small, bounded residuals, with the \(y\)-axis errors tightening more uniformly than the \(x\)-axis errors. This indicates stable behavior: agents hold their prescribed offsets and preserve obstacle clearance using an affine, solver-friendly eCBF constraint within the MPC solver.

\section{Conclusion}\label{CONCLUSION}

This paper presented a robust safety framework for high-order nonlinear multi-agent systems under bounded disturbances, integrating tube-based error feedback with eCBF-constrained DMPC. The central technical contribution is a support-function-based tightening that converts per-agent RPI tubes into explicit scalar safety margins for pairwise and agent--obstacle eCBF constraints, so that nominal feasibility of the distributed planner implies forward invariance of the safety sets for the true disturbed trajectories. Crucially, the tightened constraints remain affine in the control input, preserving the tractability of the resulting distributed quadratic programs; simulations confirm safe formation navigation in obstacle-rich environments under persistent disturbances with bounded formation error. Future work includes extending the framework to time-varying and delayed communication topologies, reducing conservatism via online data-driven tube adaptation, and incorporating higher-level task planning for fully autonomous multi-agent deployment.

\bibliography{ifacconf}             
                                                   







\appendix
\end{document}